\documentclass[aps,prd,showpacs]{revtex4}
\usepackage{epsf,epsfig,graphicx}
\begin{document}
\title{Restoration of $k_T$ factorization for
low $p_T$ hadron hadroproduction}

\author{Chun-peng Chang$^{1,2}$} \email{timpanii@gmail.com}

\author{Hsiang-nan Li$^{1,2,3,4}$} \email{hnli@phys.sinica.edu.tw}

\affiliation{$^1$Institute of Physics, Academia Sinica, Taipei,
Taiwan 115, Republic of China,}

\affiliation{$^2$Department of Physics, National Tsing-Hua
University, Hsinchu, Taiwan 300, Republic of China}

\affiliation{$^3$Department of Physics, National Cheng-Kung
University, Tainan, Taiwan 701, Republic of China}

\affiliation{$^4$Institute of Applied Physics, National Cheng-Chi
University, Taipei, Taiwan 116, Republic of China}

\begin{abstract}

We discuss the applicability of the $k_T$ factorization theorem to
low-$p_T$ hadron production in hadron-hadron collision in a simple
toy model, which involves only scalar particles and gluons. It has
been shown that the $k_T$ factorization for high-$p_T$ hadron
hadroproduction is broken by soft gluons in the Glauber region,
which are exchanged among a transverse-momentum-dependent (TMD)
parton density and other subprocesses of the collision. We explain
that the contour of a loop momentum can be deformed away from the
Glauber region at low $p_T$, so the above residual infrared
divergence is factorized by means of the standard eikonal
approximation. The $k_T$ factorization is then restored in the sense
that a TMD parton density maintains its universality. Because the
resultant Glauber factor is independent of hadron flavors,
experimental constraints on its behavior are possible. The $k_T$
factorization can also be restored for the transverse single-spin
asymmetry in hadron-hadron collision at low $p_T$ in a similar way,
with the residual infrared divergence being factorized into the same
Glauber factor.

\end{abstract}


\maketitle

\section{INTRODUCTION}

The $k_T$ factorization theorem has been widely applied to inclusive
and exclusive processes in perturbative QCD
\cite{CCH,CE,LRS,BS,LS,HS}. This theorem holds for simple processes,
such as deeply inelastic scattering (DIS) and Drell-Yan production.
Recently, it was found that the $k_T$ factorization breaks down for
complicated processes like high-$p_T$ hadron production in
hadron-hadron collision \cite{CQ07,VY07,CQ06,RM10}
\begin{eqnarray}
H_1(p_1)+H_2(p_2)\to H_3(p_3)+H_4(p_4)+X,
\end{eqnarray}
where $p_i$ denotes the momentum of the hadron $H_i$. The kinematic
region, in which the produced hadrons are almost back-to-back
azimuthally (relative to the collision axis), was analyzed in a toy
model field theory containing scalar particles and gluons under an
abelian gauge group. The $k_T$ factorization, if applicable, gives
the leading-power differential cross section \cite{CQ07,VY07,CQ06}
\begin{eqnarray}
E_3E_4 \frac{d\sigma}{ d^3{\bf p}_3 d^3{\bf p}_4 }=\sum\int
d\sigma_{i+j\to k+l}f_{i/1} f_{j/2}d_{3/k} d_{4/l},\label{fac}
\end{eqnarray}
which involves the transverse-momentum-dependent (TMD) parton
densities $f_{i/H}$ \cite{CS81,JMY05}, the fragmentation functions
$d_{H/i}$, and the parton-level differential cross section
$d\sigma_{i+j\to k+l}$. The sum over the flavors and the integral
over momenta of the partons are implicit in the above expression.

\begin{figure}[tb]
\begin{center}
\begin{tabular}{ccc}
\includegraphics[scale=0.35]{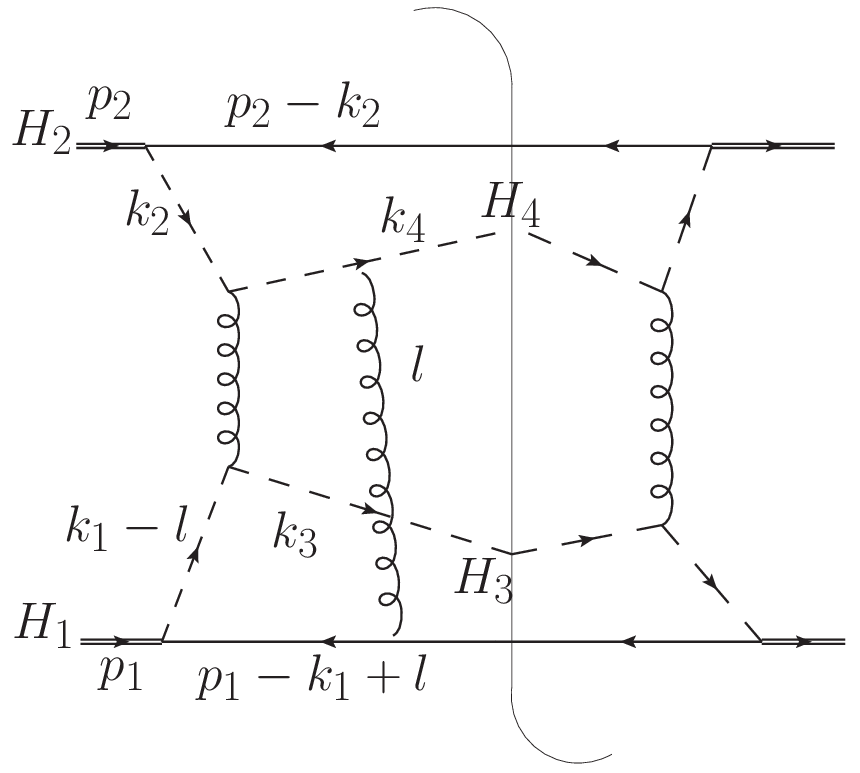}&
\includegraphics[scale=0.35]{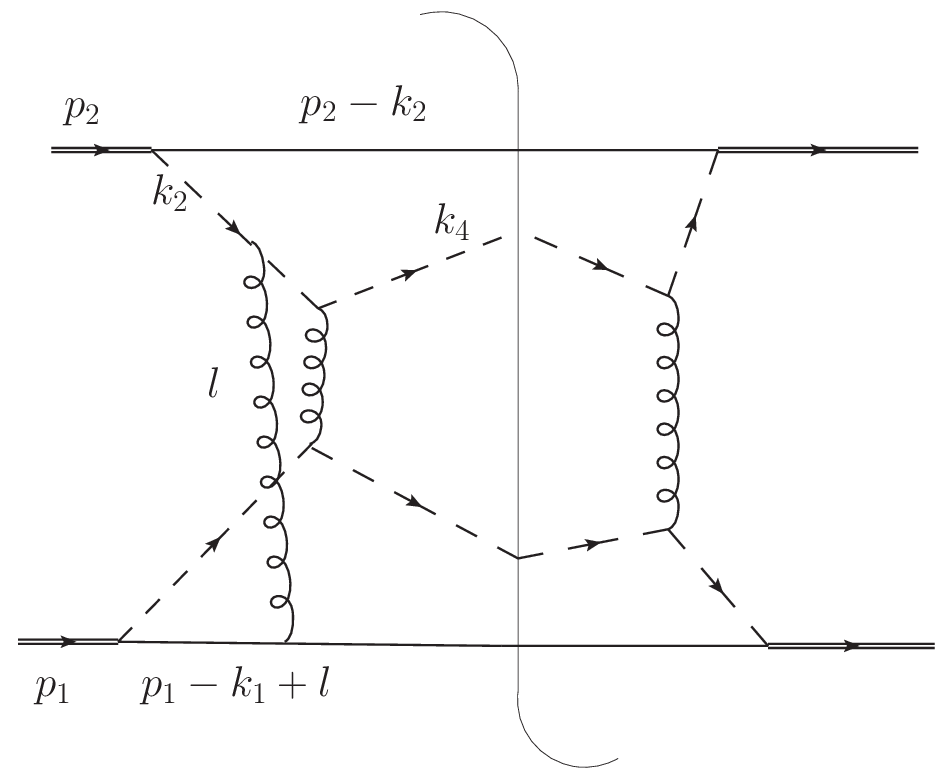}&
\includegraphics[scale=0.35]{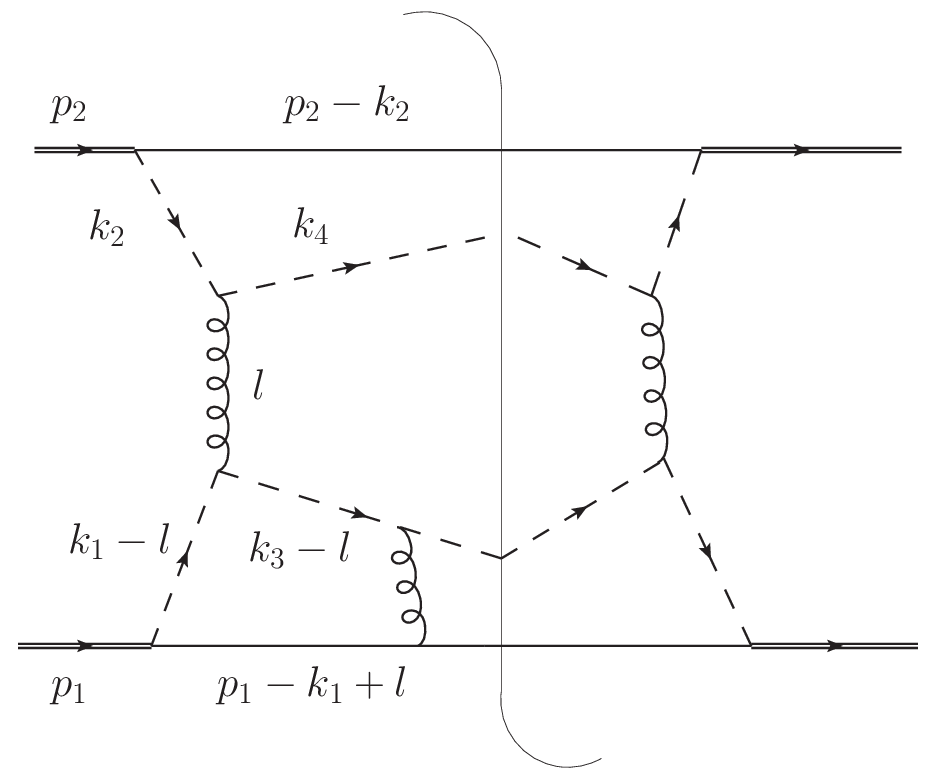}\\
(a) & (b) & (c)
\end{tabular}
\caption{Some one-loop diagrams relevant to the factorization of
$f_{i/1}$ in hadron hadroproduction. Hermitian conjugates of these
graphs also contribute. Particles in the above diagrams do not carry
colors in an abelian gauge group. In the unpolarized case, both the
solid and dashed lines represent scalars. In the case of single-spin
asymmetry, the solid lines represent fermions and the dashed lines
represent scalars.}\label{fig1}
\end{center}
\end{figure}

When factorizing the TMD parton density $f_{i/1}$, infrared
divergences from gluon exchanges among $f_{i/1}$ and other
subprocesses of the collision were identified \cite{VY07,CQ06}.
These divergences, violating the universality of $f_{i/1}$, break
the $k_T$ factorization for the hadron hadroproduction. The source
of the factorization breakdown is briefly explained below. Consider
the one-loop diagrams in Figs.~\ref{fig1}(a), \ref{fig1}(b), and
\ref{fig1}(c) with radiative gluons being emitted by the spectator
in $H_1$ and attaching to the active partons in $H_4$, $H_2$ and
$H_3$, respectively. These active parton lines can be eikonalized,
if focusing on the collection of infrared divergences from the
region with the loop momentum $l$ collinear to $p_1$. The eikonal
line from $H_3$ corresponds to the Wilson line appearing in the
operator definition of $f_{i/1}$. The collinear divergences in
Figs.~\ref{fig1}(a) and \ref{fig1}(b), which carry the information
of the TMD parton density $f_{j/2}$, need to cancel in order to
ensure the universality of $f_{i/1}$. However, the eikonal
propagators associated with the partons in $H_2$ and $H_4$ are
summed into an imaginary piece
\begin{eqnarray}
\frac{1}{-l^++i\epsilon} + \frac{1}{l^++i\epsilon}  = -2\pi i
\delta(l^+),\label{eik1}
\end{eqnarray}
where the first (second) term comes from Fig.~\ref{fig1}(a)
(\ref{fig1}(b)). The
$\delta$-function leads to a residual infrared divergence from the
Glauber region \cite{BBP81}.

This one-loop residual infrared divergence may not cause trouble, if
the leading-order (LO) amplitude ${\cal M}^{(0)}$ is real. The
expansion of the differential cross section for the hadron
hadroproduction up to next-to-leading order (NLO) gives
\begin{eqnarray}
|{\cal M}|^2 =|{\cal M}^{(0)}|^2+2{\rm Re}[{\cal M}^{(0)}{\cal
M}^{(1)*}].\label{real}
\end{eqnarray}
According to Eq.~(\ref{eik1}), the residual infrared divergence will
be purely imaginary, if ${\cal M}^{(0)}$ is real, so it does not
survive in the second term ${\rm Re}[{\cal M}^{(0)}{\cal
M}^{(1)*}]$. At two loops, the residual infrared divergence, arising
from the real product $\delta(l_1^+)\delta(l_2^+)$, does exist in
the $k_T$ factorization. This product does not break the collinear
factorization \cite{BL,ER,CZS,CZ}: it has been explicitly
demonstrated \cite{CQ06} that the two-loop residual infrared
divergence cancels, when the parton transverse momenta are
integrated out. The cancellation occurs between the diagrams with
the two Glauber gluons on the different sides of the final-state cut
and the diagrams with the two Glauber gluons on the same side.
Namely, the universality of a parton distribution function defined
for the collinear factorization is maintained in the hadron
hadroproduction. It was also found that the Glauber effect exists in
complicated dijet production in hadron-hadron collision \cite{FS09},
but does not in simple Drell-Yan processes \cite{LMa08}.

In the present work we shall have a closer look at the failure of
the $k_T$ factorization for the hadron hadroproduction. If the
residual infrared divergence could be factorized from the collision,
the universality of $f_{i/1}$ would be recovered at the price that
Eq.~(\ref{fac}) involves an additional nonperturbative input.
However, the infrared gluons are characterized by momenta in the
Glauber region as stated before, in which the spectator line of
$H_1$ can not be eikonalized \cite{CQ07,CQ06}. Therefore, the Ward
identity argument used in standard factorization proofs
\cite{GTB,CSS85} does not apply. It will be shown that the $k_T$
factorization actually holds at low $p_T$ (e.g., few GeV at
Tevatron), though it breaks down at high $p_T$ as claimed in
\cite{CQ06}. The argument is that the contour of a loop momentum can
be deformed away from the Glauber region at low $p_T$, so the
eikonalization becomes valid as a leading-power approximation. The
infrared gluons responsible for the factorization breakdown are then
factorized into a soft factor, called a Glauber factor here, and the
$k_T$ factorization is restored in the sense that $f_{i/1}$
maintains its universality. Since the Glauber factor is independent
of hadron flavors, experimental constraints on its behavior from
some processes are possible, based on which predictions for others
can be made.

In Sec.~II we analyze the pole structures of Figs.~\ref{fig1}(a) and
\ref{fig1}(b), and explain that the contour of a loop momentum can
be deformed away from the Glauber region at low $p_T$. The eikonal
approximation is then applicable to the spectator line in the hadron
$H_1$, and the residual infrared divergence discussed above is
factorizable. The factorization is performed in the impact parameter
space, and extended to two loops explicitly in Sec.~III. The
all-order operator definition of the resultant Glauber factor is
given in terms of Wilson lines. Before concluding, we show in
Sec.~IV that the $k_T$ factorization can also be restored for the
transverse single-spin asymmetry (SSA) in the low $p_T$ region of
the hadron hadroproduction, with the residual infrared divergence
being factorized into the same Glauber factor.

\section{EIKONALIZATION OF GLAUBER GLUONS}

At low $p_T$, the dominant contribution to the hadron
hadroproduction comes from the region with a small parton momentum
fraction $x_1\equiv k_1^+/p_1^+\ll 1$, so the $k_T$ factorization
theorem applies \cite{NL2}. Another example is the semi-inclusive
DIS \cite{BB08}, in which the small $x$ region is reached by
lowering the square of the transverse momentum of the outgoing
hadron with respect to the virtual photon direction. In this section
we shall explain Eq.~(\ref{eik1}), analyze the transition of the
pole structures of the loop momentum in Figs.~\ref{fig1}(a) and
\ref{fig1}(b) from large $x$ to small $x$, and then demonstrate the
eikonalization of the spectator line in $H_1$ at low $p_T$. To be
precise, we restrict ourself to the region with the hierarchy
$k_{1T} \ll k_1^+ \ll p_1^+$, $k_{1T}\sim \Lambda$ being a small
scale. This hierarchy is similar to that postulated in small-$x$
physics. The condition $\Lambda \ll k_1^+$ suppresses higher-twist
corrections to the considered process, and $k_1^+ \ll p_1^+$
justifies the eikonalization of the spectator as a leading-power
approximation. For convenience, the final-state hadrons are assumed
to be produced at central rapidity with $k_1^+\sim k_4^+\sim
k_4^-\sim k_{4T}$.

Figure~\ref{fig1}(a) contains the four denominators
\begin{eqnarray}
[(p_1-k_1+l)^2-m_q^2+i\epsilon][(k_1-l)^2-m_q^2+i\epsilon]
[(k_4-l)^2-m_q^2+i\epsilon](l^2-m_g^2+i\epsilon),\label{4a}
\end{eqnarray}
with $m_q$ and $m_g$ being the parton and gluon masses,
respectively, which define the following poles in the $l^+$ plane,
\begin{eqnarray}
l^+&=&k_1^+-p_1^++\frac{|{\bf l}_T-{\bf
k}_{1T}|^2+m_q^2}{2(l^--k_1^-)}
-i\epsilon (-i\epsilon),\label{po3}\\
l^+&=&k_1^++\frac{|{\bf l}_T-{\bf
k}_{1T}|^2+m_q^2}{2(l^--k_1^-)}-i\epsilon(-i\epsilon),
\label{po1}\\
l^+&=&\frac{2k_4^+l^--2{\bf
k}_{4T}\cdot {\bf l}_T+l_T^2}{2(l^--k_4^-)}+i\epsilon(+i\epsilon),\label{po2}\\
l^+&=&\frac{l_T^2+m_g^2}{2l^-}-i\epsilon(+i\epsilon), \label{po}
\end{eqnarray}
for the range $0<l^-<k_4^-$ ($k_1^-<l^-<0$). To get the pole in
Eq.~(\ref{po2}), we have employed the on-shell condition
$k_4^2=m_q^2$. Note that
$k_1^-=(k_{1T}^2+m_q^2)/[2(k_1^+-p_1^+)]\sim \Lambda^2/E$ is
negative from the on-shell condition $(p_1-k_1)^2=m_q^2$, with $E$
being the center-of-mass energy. There is no pinched singularity for
$l^->k_4^-$ and for $l^-<k_1^-$, because all the $l^+$ poles are
located in the same half plane. We do not consider the region with
$l$ collinear to $k_4$, which is relevant to the factorization of
the fragmentation function $d_{4/l}$. That is, $l^-$ is not of
$O(k_4^-)$. The first two poles in Eqs.~(\ref{po3}) and (\ref{po1}),
being away from the origin, do not pinch the contour of $l^+$. The
two poles in Eqs.~(\ref{po2}) and (\ref{po}), located in the
different half planes for $0<l^-<k_4^-$, are nearest to the origin
when the loop momentum components scale like $l^-\sim l_T\sim
\Lambda$. If $l^-$ is larger (smaller) than $l_T$, the pole in
Eq.~(\ref{po2}) (Eq.~(\ref{po})) will move away from the origin. It
implies that the contour of $l^+$ can only be pinched down to the
scale of $O(\Lambda)$. Hence, there are two leading infrared
regions for Fig.~\ref{fig1}(a), the collinear region with $l^+\sim
k_1^+$, and the soft region with $l^+\sim\Lambda$. The Glauber
region with $l^+\sim \Lambda^2/E$ is not pinched and not leading
here, an observation consistent with that made in \cite{CQ07}.

Figure~\ref{fig1}(b) contains the four denominators
\begin{eqnarray}
[(p_1-k_1+l)^2-m_q^2+i\epsilon][(k_1-l)^2-m_q^2+i\epsilon]
[(k_2+l)^2-m_q^2+i\epsilon](l^2-m_g^2+i\epsilon),\label{4b}
\end{eqnarray}
which define the following poles
\begin{eqnarray}
l^+&=&k_1^+-p_1^++\frac{|{\bf l}_T-{\bf
k}_{1T}|^2+m_q^2}{2(l^--k_1^-)}
+i\epsilon (-i\epsilon),\label{po4}\\
l^+&=&k_1^++\frac{|{\bf l}_T-{\bf
k}_{1T}|^2+m_q^2}{2(l^--k_1^-)}+i\epsilon(-i\epsilon),
\label{po5}\\
l^+&=&-k_2^++\frac{|{\bf l}_T+{\bf
k}_{2T}|^2+m_q^2}{2(l^-+k_2^-)}-i\epsilon(-i\epsilon),\label{po6}\\
l^+&=&\frac{l_T^2+m_g^2}{2l^-}+i\epsilon(+i\epsilon), \label{po7}
\end{eqnarray}
for the range $-k_2^-<l^-<k_1^-$ ($k_1^-<l^-<0$). Since the hadron
$H_2$ moves in the minus direction, the on-shell condition
$(p_2-k_2)^2=m_q^2$ implies
$k_2^+=(k_{2T}^2+m_q^2)/[2(k_2^--p_2^-)]\sim \Lambda^2/E$. There is
no pinched singularity for $l^->0$ and for $l^-<-k_2^-$, because all
the $l^+$ poles are located in the same half plane. We do not
consider the region with $l$ collinear to $k_2$, which is relevant
to the factorization of the TMD parton density $f_{j/2}$. That is,
$l^-$ is not of $O(-k_2^-)$. The two poles in Eqs.~(\ref{po6}) and
(\ref{po7}), located in the different half planes, then imply that
the contour of $l^+$ could remain away from the origin at least by
$O(\Lambda)$. This observation does not depend on the order of
magnitude of the other two poles. Namely, the leading regions of
$l^+$ for Fig.~\ref{fig1}(b) are also collinear and soft, as claimed
in \cite{CQ07}.

The soft divergences from the loop momentum $l^\mu=(l^+,l^-,{\bf
l}_T)\sim(\Lambda,\Lambda,\Lambda)$ are factorized into the ordinary
soft function by means of the eikonal approximation
\cite{JMY05,CRS08,LM08}. Other soft gluons, such as those exchanged
between the active partons of $H_1$ and $H_2$, are treated
similarly. In this case the parton, after emitting a soft gluon of
the momentum $l$, carries the momentum $k_1-l$. We have the
hierarchy $k_1^+l^-\gg |{\bf k}_{1T}-{\bf l}_T|^2$ in the considered
region with $k_1^+\gg k_{1T}, l_T$, which leads to the eikonal
approximation $1/(k_1-l)^2\approx 1/(-2k_1\cdot l)$. After handling
soft gluons, we can safely deform the contour of $l^+$ into the
collinear region. As a consequence, the hierarchical relation
$k_4^-l^+ \gg {\bf k}_{4T}\cdot {\bf l}_T\gg k_4^+l^-$ holds, and
the denominator $(k_4-l)^2-m_q^2+i\epsilon\approx -2k_4\cdot
l+i\epsilon\approx -2k_4^-l^++i\epsilon$ corresponds to the first
eikonal propagator in Eq.~(\ref{eik1}). Similarly, the hierarchical
relation $k_2^-l^+ \gg {\bf k}_{2T}\cdot {\bf l}_T\gg k_2^+l^-$ is
justified, and the denominator $(k_2+l)^2-m_q^2+i\epsilon\approx
2k_2^-l^++i\epsilon$ corresponds to the second eikonal propagator in
Eq.~(\ref{eik1}). Note that the above observation does not depend
on whether there exists a hierarchy between the minus components of
the momenta in $H_2$, because the $k_2$ dependence has disappeared
under the eikonalization. In summary, the integrations for
Figs.~\ref{fig1}(a) and \ref{fig1}(b) along the real axis of $l^+$
from $-\infty$ to $+\infty$ are equal to the integrations along the
deformed contour away from the origin, on which the eikonal
approximation holds. The result of the integrations along the
deformed contour can be obtained by substituting $l^+=0$ into the
integrands, according to Eq.~(\ref{eik1}).

\begin{figure}[tb]
\begin{center}
\begin{tabular}{ccc}
\includegraphics[scale=0.3]{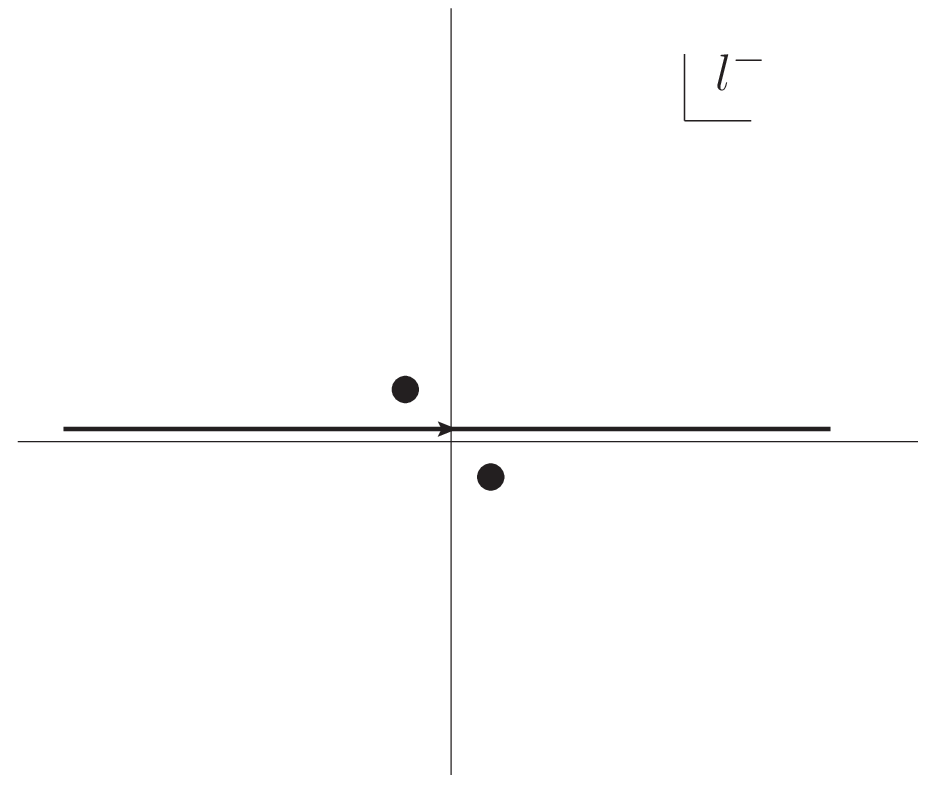}&
\includegraphics[scale=0.3]{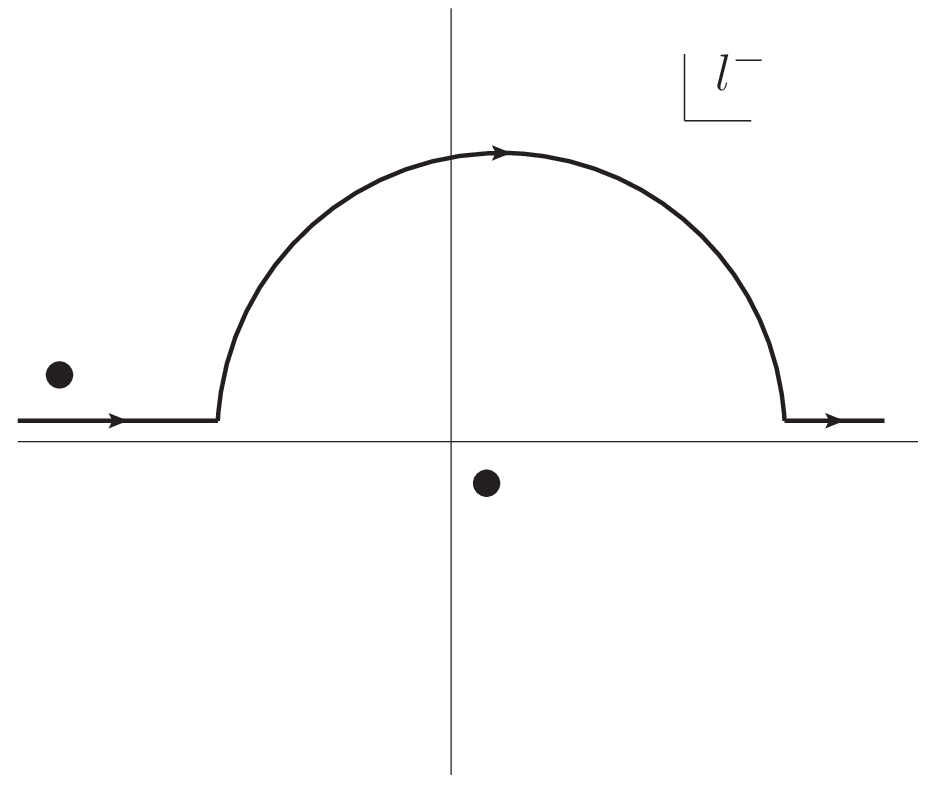}\\
(a) & (b)
\end{tabular}
\caption{Contours of $l^-$ at (a) high $p_T$ and (b) low
$p_T$.}\label{fig2}
\end{center}
\end{figure}

Following the above reasoning, the residual infrared divergence from
Figs.~\ref{fig1}(a) and \ref{fig1}(b) is collected by
\begin{eqnarray}
T_{L}^{(1)}&=&2\pi\lambda
g^4\int\frac{d^4l}{(2\pi)^4}\frac{2(p_1^+-k_1^+) (k_2+k_4)\cdot
(k_1+k_3)\delta(l^+)}{[(p_1-k_1+l)^2-m_q^2][(k_1-l)^2-m_q^2](l^2-m_g^2)
(-2k_1^+k_3^--|{\bf k}_{1T}-{\bf k}_{3T}-{\bf
l}_T|^2)}\cdots\;,\label{itab}
\end{eqnarray}
where only the relevant piece of Feynman rules
is shown explicitly. The constants $\lambda$
and $g$ denote the triple-scalar coupling and the gluon-scalar
coupling, respectively \footnote{In Refs.~\cite{CQ07,CQ06}, the
gluon-scalar coupling associated with the hadrons $H_1$ and $H_3$,
and the gluon-scalar coupling associated with $H_2$ and $H_4$ are
differentiated in order to trace the Glauber divergence. This
differentiation is in fact not necessary. In real theories, such as
QCD, there is only one coupling constant.}, and the small masses
$m_q$ and $m_g$ serve as the infrared regulators. The transverse
loop momentum $l_T$ in the numerator, being smaller than other
terms, has been dropped \cite{NL2,LM08}. Because of $l^2=-l_T^2$,
the diagrams with real gluon emissions were not included in
Fig.~\ref{fig1} \cite{CQ06}. We then consider the two poles in the
$l^-$ complex plane for Eq.~(\ref{itab}),
\begin{eqnarray}
l^-=k_1^--\frac{|{\bf l}_T-{\bf k}_{1T}|^2+m_q^2}{2k_1^+}+i\epsilon,
\;\;\;\; l^-=k_1^-+\frac{|{\bf l}_T-{\bf
k}_{1T}|^2+m_q^2}{2(p_1^+-k_1^+)} -i\epsilon\;,\label{pole}
\end{eqnarray}
with $k_1^-\sim \Lambda^2/E$. For an ordinary $x_1$ of order unity,
i.e., for $k_1^+\sim p_1^+$, these two poles in different half
planes are close to each other, such that the contour of $l^-$ is
pinched and must go through the Glauber region of $l^-\sim
\Lambda^2/E$ as shown in Fig~\ref{fig2}(a). The difficulty for
factorization caused by a pinched Glauber singularity has been
explained in \cite{CSS85}: the two terms $|{\bf l}_T-{\bf
k}_{1T}|^2$ and $(p_1^+-k_1^+)l^-$ are both of $O(\Lambda^2)$ in
this region, so the former is not negligible, and the spectator
propagator $1/[(p_1-k_1+l)^2-m_q^2]$ can not be eikonalized into
$1/[2(p_1^+-k_1^+)l^-]$. Picking up the second pole in
Eq.~(\ref{pole}), Eq.~(\ref{itab}) leads to
\begin{eqnarray}
T_{L}^{(1)}=-4\pi^2 i\lambda g^4\int\frac{d^2l_T}{(2\pi)^4}
\frac{1-x_1} {(|{\bf k}_{1T}-{\bf l}_T|^2+m_q^2)(l_T^2+m_g^2)}\frac{
(k_2+k_4)\cdot(k_1+k_3)}{-2k_1^+k_3^--|{\bf k}_{1T}-{\bf
k}_{3T}-{\bf l}_T|^2}\cdots. \label{lt}
\end{eqnarray}

For a small $x_1$ or $k_1^+\ll p_1^+$, the first pole moves away
form the origin and becomes located at $l^-\sim \Lambda^2/k_1^+$,
while the second one remains of $O(\Lambda^2/E)$. One can then
deform the contour of $l^-$ in the complex plane, so that the region
of $l^-\sim \Lambda^2/E$ is avoided as shown in Fig.~\ref{fig2}(b).
We thus have the hierarchy $(p_1^+-k_1^+)l^-\sim
(p_1^+/k_1^+)\Lambda^2\gg  |{\bf k}_{1T}-{\bf l}_T|^2\sim
\Lambda^2$, and the eikonal approximation is justified for the
spectator line in $H_1$. To verify our argument, we compare the
result from Eq.~(\ref{itab}) and that from the simplified integral
with the eikonal approximation, $2(p_1^+-k_1^+)/(p_1-k_1+l)^2\approx
1/l^-$,
\begin{eqnarray}
T_{L}^{(1)eik}&=&2\pi\lambda g^4\int\frac{d^4l}{(2\pi)^4}\frac{
(k_2+k_4)\cdot(k_1+k_3)\delta(l^+)}{[(k_1-l)^2-m_q^2] l^-(l^2-m_g^2)
(-2k_1^+k_3^--|{\bf k}_{1T}-{\bf k}_{3T}-{\bf l}_T|^2)}\cdots
.\label{eikt}
\end{eqnarray}
The pole $l^-=0-i\epsilon$ from the eikonal propagator,
corresponding to the second pole in Eq.~(\ref{pole}), gives
\begin{eqnarray}
T_{L}^{(1)eik}=-4\pi^2 i\lambda g^4\int\frac{d^2l_T}{(2\pi)^4}
\frac{1} {(|{\bf k}_{1T}-{\bf
l}_T|^2-2k_1^+k_1^-+m_q^2)(l_T^2+m_g^2)}
\frac{(k_2+k_4)\cdot(k_1+k_3)}{-2k_1^+k_3^--|{\bf k}_{1T}-{\bf
k}_{3T}-{\bf l}_T|^2} \cdots,\label{ltp}
\end{eqnarray}
where $2k_1^+k_1^-$ can be reexpressed as $x_1k_{1T}^2/(x_1-1)$
after employing $(p_1-k_1)^2=m_q^2$. It is obvious that
Eqs.~(\ref{lt}) and (\ref{ltp}) are quite different in the large
$x_1\sim O(1)$ region, but identical, if the terms proportional to
$x_1$ are neglected.

We work out the $l_T$ integrations in Eqs.~(\ref{lt}) and
(\ref{ltp}) explicitly, ignoring the transverse momentum
dependence in the hard kernel for simplicity. The expressions
\begin{eqnarray}
T_{L}^{(1)}&\approx&i\frac{\lambda
g^4}{4\pi}\left(\frac{1-x_1}{k_{1T}^2+m_q^2}\ln\frac{k_{1T}^2}
{m_g^2}+\frac{1-x_1}{k_{1T}^2+m_g^2}\ln\frac{k_{1T}^2}
{m_q^2}\right)\frac{ (k_2+k_4)\cdot(k_1+k_3)}{2k_1^+k_3^-}\cdots,
\label{lt2}\\
T_{L}^{(1)eik}&\approx&i\frac{\lambda
g^4}{4\pi}\left[\frac{1-x_1}{k_{1T}^2+(1-x_1)m_q^2}\ln\frac{k_{1T}^2}
{(1-x_1)m_g^2}+\frac{1}{k_{1T}^2+m_g^2}\ln\frac{k_{1T}^2}
{(1-x_1)m_q^2+x_1k_{1T}^2}\right]\nonumber\\
& &\times\frac{
(k_2+k_4)\cdot(k_1+k_3)}{2k_1^+k_3^-}\cdots.\label{ltp2}
\end{eqnarray}
indeed contain the same infrared logarithms, when the terms
proportional to $x_1$ are dropped. That is, Eq.~(\ref{ltp2}) under
the eikonal approximation collects the infrared logarithms in the
original loop integral, Eq.~(\ref{lt2}), at leading power of small
$x_1$. It confirms the above argument for the eikonal approximation
based on the contour deformation at low $p_T$.

We investigate whether the condition allowing the contour
deformation would be lost, when higher-order corrections to the TMD
parton density $f_{i/1}$ are taken into account. For example, the
second radiative gluon can be added between the spectator and the
outgoing parton of the momentum $k_3$. This type of gluon exchanges
may be collinear and redistribute the momenta between the spectator
and the active parton of $H_1$, such that the separation of the two
poles in Eq.~(\ref{pole}) is reduced. Assume that the active parton
carries the momentum $k_1-l_1-l_2$, and that the spectator carries
the momentum $p_1-k_1+l_1+l_2$ before emitting the second gluon, and
$p_1-k_1+l_1$ before emitting the first. Performing the contour
integration over $l_2^-$, we find the range
$-(p_1^+-k_1^+)<l_2^+<k_1^+$ for the existence of pinched
singularities. The poles of $l_1^-$ in Eq.~(\ref{pole}) are modified
into
\begin{eqnarray}
l_1^-=k_1^--l_2^--\frac{|{\bf l}_{1T}+{\bf l}_{2T}-{\bf
k}_{1T}|^2+m_q^2} {2(k_1^+-l_2^+)}+i\epsilon,\;\;\;\;
l_1^-=k_1^--l_2^-+\frac{|{\bf l}_{1T}+{\bf l}_{2T}-{\bf
k}_{1T}|^2+m_q^2}{2(p_1^+-k_1^++l_2^+)} -i\epsilon\;,\label{pole3}
\end{eqnarray}
where the value of $l_2^-$ is determined by its pole enclosed by the
contour of $l_2^-$. The above two poles are far apart from each
other as long as $l_2^+$ is of $O(k_1^+)$ or smaller, for which the
contour of $l_1^-$ can be deformed away from the Glauber region. One
may worry that they become close to each other as $l_2^+$ reaches
$O(-p_1^+)$. In this case, we have $l_2^-\sim\Lambda^2/E$ determined
by the contour integration over $l_2^-$, and the Glauber region of
$l_1^-$ is pinched. However, the scattered particle, with the
invariant mass squared $(k_3-l_2)^2\sim -2k_3^-l_2^+$, will be more
off-shell in the latter case than in the former case. That is, the
contribution from the latter is suppressed by a power of $x_1\equiv
k_1^+/p_1^+$, and should be neglected. We conclude that the eikonal
approximation for the spectator line of $H_1$ holds at leading power
of $x_1$, even when including higher-order corrections to $f_{i/1}$.

It has been pointed out that the naive definition for a TMD parton
density with light-like Wilson lines develops light-cone
singularities from the region with a loop momentum collinear to
Wilson lines \cite{Co03}. Two methods have been proposed to
regularize these light-cone singularities in \cite{Co03}. Because we
work on light-like Wilson lines as indicated in Eq.~(\ref{eik1}), it
could be understood that we have implicitly adopted the method with
a subtraction factor given by matrix elements of four Wilson lines
\cite{FH07}. This subtraction factor is constructed by means of
eikonal approximation for diagrams which contribute to a TMD parton
density, such as Fig.~\ref{fig1}(c). For more details of the
construction, refer to \cite{CH00}. Since the subtraction factor
comes from the eikonalization of a TMD parton density, it is not
involved in the discussion of the factorization breaking effects
here.

Another method to regularize light-cone singularities is to rotate
the light-like Wilson lines for a TMD parton density away from
the light cone \cite{Co03,NL07}. It is then worth examining the
effect of replacing the Wilson line direction $u_-^\mu=(0,1,{\bf
0}_T)$ by $n^\mu=(n^+,n^-,{\bf 0}_T)$ with $n^2<0$. The three
denominators in Eq.~(\ref{itab}), containing the terms
$2(l^+-k_1^+)l^-$, $2(l^++p_1^+-k_1^+)l^-$, and $2l^+l^-$, lead to
three poles in different $l^-$ half planes for the range $0> l^+ >
-(p_1^+-k_1^+)$,
\begin{eqnarray}
& &l^-=k_1^-+\frac{|{\bf l}_T-{\bf k}_{1T}|^2}{2(l^+-k_1^+)}
+i\epsilon,\label{p1}\\
& &l^-=k_1^-+\frac{|{\bf l}_T-{\bf k}_{1T}|^2}{2(l^++p_1^+-k_1^+)}
-i\epsilon,\label{p2}\\
& &l^-=\frac{l_T^2}{2l^+}+i\epsilon.\label{p3}
\end{eqnarray}
The quark and gluon masses were not shown explicitly in the above
expressions for simplicity.
Solving Eqs.~(\ref{p1}), (\ref{p2}), and (\ref{p3}) together with
the constraint $n^+l^-+n^-l^+=0$ from the $\delta$-function
$\delta(n\cdot l)$, we obtain
\begin{eqnarray}
l^+&=&\frac{1}{2}\left[k_1^+-k_1^-\frac{n^+}{n^-}-
\sqrt{\left(k_1^++k_1^-\frac{n^+}{n^-}\right)^2
-2\frac{n^+}{n^-}|{\bf l}_T-{\bf k}_{1T}|^2}\right],\label{s1}\\
l^+&=&\frac{1}{2}\left[k_1^+-p_1^+-k_1^-\frac{n^+}{n^-}+
\sqrt{\left(p_1^+-k_1^+-k_1^-\frac{n^+}{n^-}\right)^2
-2\frac{n^+}{n^-}|{\bf l}_T-{\bf k}_{1T}|^2}\right],\label{s2}\\
l^+&=&-\sqrt{\frac{-n^+}{2n^-}}l_T,\label{s3}
\end{eqnarray}
respectively. It is obvious that the choice $n^2<0$ prevents complex
solutions. We have checked that no solution of $l^+$ exists in the
range $k_1^+ > l^+ >0$, for which the three poles of $l^-$ are also
located in different half planes. Equations~(\ref{p1}) and
(\ref{s1}) imply the pole $l^-\sim \Lambda^2/k_1^++i\epsilon$, if
$n^+$ and $n^-$ are of the same order of magnitude.
Equations~(\ref{p2}) and (\ref{s2}) [(\ref{p3}) and (\ref{s3})]
imply the pole $l^-\sim \Lambda^2/E-i\epsilon$ ($l^-\sim
\Lambda+i\epsilon$). Hence, the contour of $l^-$ can be deformed,
such that $l^-$ remains of $O(\Lambda^2/k_1^+)$ in the contour
integration, and the eikonal approximation for the spectator
line of $H_1$ is justified. That is, the conclusion drawn in this
section holds even under the variation of the Wilson
line direction.

\section{FACTORIZATION IN IMPACT PARAMETER SPACE}

In this section we shall sum the residual infrared divergences from
the Glauber gluons to all orders in the impact parameter space. The
factors $\lambda/(|{\bf k}_{1T}-{\bf l}_T|^2+m_q^2)$ and
$g^2(k_2+k_4)\cdot(k_1+k_3)/(-2k_1^+k_3^--|{\bf k}_{1T}-{\bf
k}_{3T}-{\bf l}_T|^2)$ in Eq.~(\ref{ltp}) are absorbed into the LO
TMD parton density $f_{i/1}^{(0)}$ and the LO parton-level
differential cross section $d\sigma_{i+j\to k+l}^{(0)}$,
respectively. The contribution from Figs.~\ref{fig1}(a) and
\ref{fig1}(b) is then factorized into the convolution
\begin{eqnarray}
T_{L}^{(1)}\approx-i\frac{g^2}{(2\pi)^2}\int
\frac{d^2l_T}{l_T^2+m_g^2}d\sigma_{i+j\to k+l}^{(0)}({\bf
k}_{1T}-{\bf l}_T-{\bf k}_{3T},{\bf k}_{1T}-{\bf k}_{3T})
f_{i/1}^{(0)}({\bf k}_{1T}-{\bf l}_T,{\bf k}_{1T})\cdots,\label{ts1}
\end{eqnarray}
where the soft divergence from $l_T\to 0$ is apparent. The argument
${\bf k}_{1T}-{\bf l}_T$ (${\bf k}_{1T}$) in $f_{i/1}^{(0)}$ labels
the parton transverse momentum in the hadron $H_1$ before (after)
the final-state cut. The two arguments in $d\sigma_{i+j\to
k+l}^{(0)}$ indicate that two active partons from $H_1$ participate
the hard scattering actually: one parton corresponds to the valence
scalar particle of $H_1$, and another to the Glauber gluon.

Equation~(\ref{ts1}) is rewritten, in the impact parameter space, as
\begin{eqnarray}
T_{L}^{(1)}&\approx&\int d^2b_l d^2b_rd^2b'_ld^2b'_r [-iS({\bf
b}_l)]d\hat\sigma^{(0)}_{i+j\to k+l}({\bf b}_l-{\bf b}'_l,{\bf
b}_r-{\bf b}'_r)\hat f_{i/1}^{(0)}({\bf b}'_l,{\bf b}'_r)\nonumber\\
& &\times e^{i{\bf k}_{1T}\cdot({\bf b}_l-{\bf b}_r)}e^{-i{\bf
k}_{3T}\cdot({\bf b}_l-{\bf b}_r-{\bf b}'_l+{\bf
b}'_r)}\cdots,\label{left}
\end{eqnarray}
with the one-loop Glauber factor
\begin{eqnarray}
S({\bf b})=\frac{g^2}{(2\pi)^2} \int
\frac{d^2l_T}{l_T^2+m_g^2}e^{-i{\bf l}_T\cdot {\bf
b}}=\frac{g^2}{2\pi} K_0(bm_g),\label{sf}
\end{eqnarray}
where $K_0$ is the modified Bessel function.
The factor $-i$ has been made explicit, so that the Glauber factor
$S$ defined above is real. The LO TMD parton density,
arising from the separate Fourier transformations of the parton
propagators $\lambda/(k_{1T}^2+m_q^2)$ before and after the
final-state cut, is given by
\begin{eqnarray}
\hat f_{i/1}^{(0)}({\bf b}'_l,{\bf b}'_r)=
\frac{\lambda^2}{4\pi^2}K_0(b'_lm_q)K_0(b'_rm_q).\label{LO}
\end{eqnarray}

If the Glauber gluons appear on the right-hand side of the
final-state cut, we have
\begin{eqnarray}
T_{R}^{(1)}&\approx&\int d^2b_l d^2b_rd^2b'_ld^2b'_r
d\hat\sigma^{(0)}_{i+j\to k+l}({\bf b}_l-{\bf b}'_l,{\bf b}_r-{\bf
b}'_r) [iS({\bf b}_r)]\hat f_{i/1}^{(0)}({\bf b}'_l,{\bf b}'_r)\nonumber\\
& &\times e^{i{\bf k}_{1T}\cdot({\bf b}_l-{\bf b}_r)}e^{-i{\bf
k}_{3T}\cdot({\bf b}_l-{\bf b}_r-{\bf b}'_l+{\bf
b}'_r)}\cdots.\label{right}
\end{eqnarray}
It is easy to see that Eq.~(\ref{left}) and Eq.~(\ref{right}) cancel
each other, because $d\hat\sigma_{i+j\to k+l}$ and $\hat f_{i/1}$
are symmetric under the interchange of their two arguments. This
cancellation verifies that the Glauber divergence does not
cause a problem at one loop, as stated after Eq.~(\ref{real}). Here
we give more explanation to the notations adopted in the above
expressions, which correspond to the choice of the triple-scalar
vertex in $H_1$ as the origin of the transverse coordinates. Choices
of other vertices as the origin are certainly allowed. The variables
${\bf b}'_l$ and ${\bf b}'_r$ denote the transverse coordinates of
the partons coming out of $H_1$, namely, the lower ends of the hard
gluons, before and after the final-state cut, respectively. ${\bf
b}_l$ and ${\bf b}_r$ denote the transverse coordinates of the upper
ends of the hard gluons before and after the final-state cut,
respectively. The Bessel function $K_0$ in Eq.~(\ref{sf}) then
describes the gluon propagation in the transverse plane from the
triple-scalar vertex to the upper end of the hard gluon. The
propagation in the impact parameter space in terms of the Bessel
function $K_0$ has been also obtained in \cite{FH07}.

At next-to-next-to-leading (NNLO) order, one more Glauber gluon
attaches to the spectator line of $H_1$ in Fig.~\ref{fig1}. We have
either one Glauber gluon on each side of the final-state cut or
two Glauber gluons on the same side. The former case can be
handled by repeating the analysis in the previous section, giving
\begin{eqnarray}
T_{LR}^{(2)}&\approx&\int d^2b_l d^2b_rd^2b'_ld^2b'_r[-iS({\bf
b}_l)] d\hat\sigma^{(0)}_{i+j\to k+l}({\bf b}_l-{\bf b}'_l,{\bf
b}_r-{\bf b}'_r)[iS({\bf b}_r)]
\hat f_{i/1}^{(0)}({\bf b}'_l,{\bf b}'_r)\nonumber\\
& &\times e^{i{\bf k}_{1T}\cdot({\bf b}_l-{\bf b}_r)}e^{-i{\bf
k}_{3T}\cdot({\bf b}_l-{\bf b}_r-{\bf b}'_l+{\bf b}'_r)}\cdots.
\end{eqnarray}
For the latter with the two Glauber gluons on the left-hand side,
we assume that the active parton carries the momentum $k_1-l_1-l_2$,
and that the spectator carries the momentum $p_1-k_1+l_1+l_2$ before
emitting the second gluon, and $p_1-k_1+l_1$ before emitting the
first. According to the explanation in the previous section, the
plus components of the Glauber gluon momenta vanish due to the
associated product $\delta(l_1^+) \delta(l_2^+)$ \cite{CQ06}. We
then consider the poles from the first two propagators in the
$l^{\prime -}\equiv l_1^-+l_2^-$ complex plane:
\begin{eqnarray}
l^{\prime -}=k_1^--\frac{|{\bf l}_{1T}+{\bf l}_{2T}-{\bf k}_{1T}|^2}
{2k_1^+}+i\epsilon,\;\;\;\; l^{\prime -}=k_1^-+\frac{|{\bf
l}_{1T}+{\bf l}_{2T}-{\bf k}_{1T}|^2}{2(p_1^+-k_1^+)}
-i\epsilon\;,\label{pole2}
\end{eqnarray}
which are apart from each other by $O(\Lambda^2/k_1^+)$. Similarly,
the contour of $l^{\prime -}$ can be deformed to avoid the Glauber
region, such that the spectator propagator $1/(p_1-k_1+l_1+l_2)^2$
is eikonalized into $1/(l_1^-+l_2^-+i\epsilon)$. The last propagator
$1/(p_1-k_1+l_1)^2$, with a single pole in the $l_1^-$ plane, is
eikonalized into $1/(l_1^-+i\epsilon)$.

Exchanging the gluon of the momentum $l_1$ and the gluon of $l_2$,
the resultant diagram contains the two eikonal
propagators $1/(l_1^-+l_2^-+i\epsilon)$ and $1/(l_2^-+i\epsilon)$.
The sum of these two NNLO diagrams obeys the relation
\cite{CQ06}
\begin{eqnarray}
\frac{1}{(l_1^-+l_2^-)+i\epsilon}\frac{1}{l_1^-+i\epsilon}
+\frac{1}{(l_1^-+l_2^-)+i\epsilon}\frac{1}{l_2^-+i\epsilon}
=\frac{1}{l_1^-+i\epsilon}\frac{1}{l_2^-+i\epsilon},\label{eiko}
\end{eqnarray}
which is crucial for the factorization of the Glauber gluons from
the process. It will not hold, if the transverse loop momenta are
retained in the denominators. Performing the contour integrations
over $l_1^-$ and $l_2^-$, we derive the factorization of the Glauber
divergence
\begin{eqnarray}
T_{LL}^{(2)}&\approx&\frac{1}{2}\left[-i\frac{g^2}{(2\pi)^2}
\right]^2\int \frac{d^2l_{1T}}{l_{1T}^2}\int
\frac{d^2l_{2T}}{l_{2T}^2}d\sigma_{i+j\to k+l}^{(0)}({\bf
k}_{1T}-{\bf l}_{1T}-{\bf l}_{2T}-{\bf k}_{3T},
{\bf k}_{1T}-{\bf k}_{3T})\nonumber\\
& &\times f_{i/1}^{(0)}({\bf k}_{1T}-{\bf l}_{1T}-{\bf l}_{2T},{\bf
k}_{1T})\cdots,\label{2loop}
\end{eqnarray}
which can be rewritten, in the impact parameter space, as
\begin{eqnarray}
T_{LL}^{(2)}&\approx&\int d^2b_ld^2b_rd^2b'_ld^2b'_r
\frac{1}{2}[-iS({\bf b}_l)]^2d\hat\sigma^{(0)}_{i+j\to k+l}({\bf
b}_l-{\bf b}'_l,{\bf b}_r-{\bf b}'_r)\hat f_{i/1}^{(0)}({\bf b}'_l,{\bf b}'_r)\nonumber\\
& &\times e^{i{\bf k}_{1T}\cdot({\bf b}_l-{\bf b}_r)}e^{-i{\bf
k}_{3T}\cdot({\bf b}_l-{\bf b}_r-{\bf b}'_l+{\bf
b}'_r)}\cdots.\label{2b}
\end{eqnarray}

Viewing Eqs.~(\ref{2loop}) and (\ref{2b}), it is clear why the
all-order summation of the Glauber divergences can be facilitated in
the $b$ convolution, instead of in the $k_T$ convolution. Because of
the simple abelian gauge interaction considered here, the
application of the above procedure to higher loops is trivial. The
diagrams with gluons being emitted by the spectator of $H_1$ and
attaching to the active partons of $H_2$ and $H_4$ are then summed
into
\begin{eqnarray}
T&\approx&\int d^2b_ld^2b_rd^2b'_ld^2b'_r
e^{-iS({\bf b}_l)}d\hat\sigma_{i+j\to k+l}({\bf b}_l-{\bf b}'_l,{\bf
b}_r-{\bf b}'_r)e^{iS({\bf b}_r)}\hat f_{i/1}({\bf b}'_l,{\bf b}'_r)\nonumber\\
& &\times e^{i{\bf k}_{1T}\cdot({\bf b}_l-{\bf b}_r)}e^{-i{\bf
k}_{3T}\cdot({\bf b}_l-{\bf b}_r-{\bf b}'_l+{\bf
b}'_r)}\cdots.\label{sig1}
\end{eqnarray}
The collinear gluon exchanges of the type in Fig.~\ref{fig1}(c) can
be factorized in the standard way, so both $d\hat\sigma_{i+j\to
k+l}$ and $\hat f_{i/1}$ have been extended to all orders. In the
small-$x$ region the spectator line of $H_1$ can always be
eikonalized according to the contour deformation, such that its
propagators do not depend on transverse momenta. Given a TMD parton
density, it is then possible to perform separate Fourier
transformations for the parton transverse momenta before and after
the final-state cut. The definition of $\hat f_{i/1}$ is related to
the usual matrix element of the nonlocal operator via
\begin{equation}
\int d^2b'\hat f_{i/1}({\bf b}',{\bf b}+{\bf b}')=\int\frac{dy^-}{2\pi}e^{-ix_1
p_1^+y^-} \langle H_1|\phi_i^\dagger(y^-,{\bf b})W_-(y^-,{\bf
b};\infty)^{\dag}W_-(0,{\bf 0};\infty)\phi_i(0,{\bf
0})|H_1\rangle,\label{dep}
\end{equation}
where the dependence on the small momentum fraction $x_1$ has been
suppressed on the left-hand side of the above expression. The factor
$W_-$ denotes the Wilson line operator
\begin{eqnarray}
\label{eq:WL.def} W_-(y^-,{\bf b};\infty) = P \exp\left[-ig
\int_0^\infty d\lambda u_-\cdot A(y+\lambda u_-)\right]\;,
\end{eqnarray}
with the gluon field $A$ and the coordinate $y=(0,y^-,{\bf b})$. It
should be understood that the two Wilson lines $W_-(y^-,{\bf
b};\infty)^{\dag}$ and $W_-(0,{\bf 0};\infty)$ are connected by a
vertical link at infinity \cite{BJY,CS08}, which does not contribute
in a covariant gauge. When the final-state hadron pair carries a net
large transverse momentum, the $k_{1T}$ dependence in the
fragmentation functions is negligible. Integrating over $k_{1T}$,
the $\delta$-function $\delta({\bf b}_l-{\bf b}_r)$ renders the soft
factor vanish, and Eq.~(\ref{sig1}) reduces to a formula in the
collinear factorization \cite{NQS05}.


The operator definition of the Glauber factor is given by
\begin{eqnarray}
e^{-iS({\bf b})}=\langle 0|W_-(0,{\bf b};-\infty)^{\dag}W_-(0,{\bf
b};\infty) W_+(0,{\bf 0};\infty)W_+(0,{\bf
0};-\infty)^{\dag}|0\rangle,\label{soft}
\end{eqnarray}
where $W_+$ denotes another Wilson line operator
\begin{eqnarray}
W_+(y^+,{\bf b};\infty) = P \exp\left[-ig \int_0^\infty d\lambda
u_+\cdot A(y+\lambda u_+)\right]\;,
\end{eqnarray}
with the coordinate $y=(y^+,0,{\bf b})$ and the dimensionless vector
$u_+^\mu=(1,0,{\bf 0}_T)$. Similarly, there exist also vertical
links among the above four Wilson lines at infinity. The
construction of Eq.~(\ref{soft}) is similar to that of
the subtraction factor in \cite{FH07,CH00}. The net effects
of $W_-(0,{\bf b};-\infty)^{\dag}W_-(0,{\bf b};\infty)$ and of
$W_+(0,{\bf 0};\infty)W_+(0,{\bf 0};-\infty)^{\dag}$ demand the
vanishing of the components $l^+$ and $l^-$ of a loop momentum,
respectively. A Glauber gluon is then off-shell by $l_T^2$ as
indicated in Eq.~(\ref{sf}). It can be shown, by expanding the
Wilson line operators order by order, that Eq.~(\ref{soft})
reproduces the Feynman rules for the Glauber factor.

To derive the differential cross section, we integrate the momentum
conservations $\delta(k_1^+-k_3^+-k_4^+)$,
$\delta(k_2^--k_3^--k_4^-)$, and $\delta^2({\bf k}_{1T}+{\bf
k}_{2T}-{\bf k}_{3T}-{\bf k}_{4T})$ over $k_1^+$, $k_2^-$ and
$k_{1T}$, respectively, and the on-shell conditions $\delta(k_3^2)$
and $\delta(k_4^2)$ over $k_3^0$ and $k_4^0$, respectively. The
longitudinal parton momenta $k_1^+$ and $k_2^-$ are then related to
${\bf k}_3$ and ${\bf k}_4$. Substituting ${\bf k}_{1T}={\bf
k}_{3T}+{\bf k}_{4T}-{\bf k}_{2T}$ into the Fourier factor $e^{i{\bf
k}_{1T}\cdot({\bf b}_l-{\bf b}_r)}$, and integrating over $k_{2T}$,
$f_{j/2}({\bf k}_{2T})$ is transformed into the impact parameter
space. At last, we arrive at the factorization formula modified by
the Glauber factor associated with the hadron $H_1$
\begin{eqnarray}
E_3E_4 \frac{d\sigma}{ d^3{\bf p}_3 d^3{\bf p}_4 }&=&\sum\int
\frac{d^3{\bf k}_{3}}{|{\bf k}_3|}\frac{d^3{\bf k}_{4}}{|{\bf
k}_4|}d^2b_ld^2b_rd^2b'_ld^2b'_r e^{-iS({\bf
b}_l)}d\hat\sigma_{i+j\to k+l}({\bf b}_l-{\bf b}'_l,{\bf b}_r-{\bf
b}'_r)e^{iS({\bf b}_r)}\nonumber\\
& &\hspace{1.0cm}\times \hat f_{i/1}({\bf b}'_l,{\bf b}'_r)\hat
f_{j/2}({\bf b}_l-{\bf b}_r)d_{k/3}({\bf k}_{3})d_{l/4}({\bf
k}_{4})e^{i{\bf k}_{3T}\cdot({\bf b}'_l-{\bf b}'_r)}e^{i{\bf
k}_{4T}\cdot({\bf b}_l-{\bf b}_r)},\label{ys}
\end{eqnarray}
where the dependence on ${\bf k}_3$ and ${\bf k}_4$ in
$d\hat\sigma_{i+j\to k+l}$, $\hat f_{i/1}$ and $\hat f_{j/2}$ is
implicit. The Glauber divergence associated with the hadron $H_2$
can be analyzed in the same way, which is not discussed in this
work. To confirm that the Glauber effect vanishes at one loop,
we expand the two Glauber factors in Eq.~(\ref{ys}) into
$-iS({\bf b}_l)$ and $iS({\bf b}_r)$. Performing the variable exchanges
$b_l\leftrightarrow b_r$ and $b'_l\leftrightarrow b'_r$, and
employing the symmetry under the exchange of the two arguments
of $d\hat\sigma_{i+j\to k+l}$ and of $\hat f_{i/1}$, it is easy to
see that the formula with $-iS({\bf b}_l)$ becomes identical to
the formula with $iS({\bf b}_r)$, but is opposite in sign. Namely, they
cancel each other, and the Glauber effect indeed starts
from two loops \cite{CQ07,VY07,CQ06,RM10}. Expanding the two Glauber
factors to higher orders, we verify that Eq.~(\ref{ys}) gives a real
contribution to the differential cross section.

Note that the Glauber factor in Eq.~(\ref{soft}) does not carry the
flavor indices $i$, $j$, $k$, and $l$, and is independent of the
species of hadrons involved in the collision at leading power. This
universality makes possible experimental constraints on its behavior
from some processes (e.g., $HH\to\pi\pi+X$), and predictions from
the $k_T$ factorization for other processes (e.g., $HH\to KK+X$). We
can study the Glauber effect by comparing results from
Eq.~(\ref{ys}) and from the corresponding formula without the
Glauber factor. It is emphasized that the Glauber factor differs
from the soft function obtained in the simple Drell-Yan process
\cite{CS81}, for which the $k_T$ factorization has been justified
\cite{GTB,CSS85}. The infrared divergences studied in \cite{CS81}
arise from the ordinary (not Glauber) soft region, and can always be
collected by means of the eikonalzation as explained in Sec.~II. The
soft function appears in the $k_T$ factorization for the Drell-Yan
process, because of the incomplete infrared cancellation between
virtual corrections, where loop momenta do not flow through hard
scattering, and real corrections, where loop momenta do.

\section{FACTORIZATION FOR SINGLE-SPIN ASYMMETRY}

We then investigate the applicability of the $k_T$ factorization
theorem to the transverse SSA in hadron-hadron collision
\cite{VY07,BBMP,QVY,RT07,AB07,DM07} with $H_1$ being the
transversely polarized hadron. For the SSA, the parton transverse
momentum must be taken into account, and the imaginary part of the
polarized TMD parton density contributes. Adopting a similar model
field theory \cite{BHS02}, which contains additional fermion fields,
the $k_T$ factorization for the SSA was also shown to fail
\cite{CQ07}. The mechanism is attributed to the Glauber gluons,
identical to that in the unpolarized hadron hadroproduction. This is
the reason why the definitions of the TMD parton densities in the
SSA and unpolarized processes were modified by including the same
additional Wilson links in \cite{VY07}. The sum of these additional
Wilson lines leads to the $\delta$-function in Eq.~(\ref{eik1}),
which breaks the $k_T$ factorization at one loop. In this section we
shall show that the $k_T$ factorization is restored for the SSA at
low $p_T$, where the contour of a loop momentum can be deformed away
from the Glauber region.

The loop integral associated with the sum of Figs.~\ref{fig1}(a) and
\ref{fig1}(b) for the SSA is written as
\begin{eqnarray}
S_{L}^{(1)}&=&2\pi\lambda
g^4\int\frac{d^4l}{(2\pi)^4}\frac{(k_2+k_4)\cdot
(k_1+k_3)\delta(l^+)}{[(k_1-l)^2-m_q^2] [(p_1-k_1+l)^2-m_q^2]
(l^2-m_g^2)(-2k_1^+k_3^--|{\bf k}_{1T}-{\bf
k}_{3T}-{\bf l}_T|^2)},\nonumber\\
& &\times \frac{1}{2}Tr[(\not p_1+m_H)\gamma_5\not s(\not p_1-\not
k_1+\not l+m_q)\gamma^+(\not p_1-\not k_1+m_q)],\label{ssa1}
\end{eqnarray}
in which $m_H$ denotes the mass of the hadron $H_1$, and the spin
vector $s$ is chosen in the transverse direction. The trace in
Eq.~(\ref{ssa1}) gives, at small $k_1$,
\begin{eqnarray}
\frac{1}{2}Tr[(\not p_1+m_H)\gamma_5\not s(\not p_1-\not k_1+\not
l+m_q)\gamma^+(\not p_1-\not k_1+m_q)] \approx
2i(m_H+m_q)\epsilon_{\alpha\beta}s^\alpha l^\beta
p_1^+,\label{trace}
\end{eqnarray}
where the $\epsilon$ tensor obeys $\epsilon_{12}=-\epsilon_{21}=1$,
and $l^\beta$ picks up the transverse components. Following the
similar reasoning, we eikonalize the spectator fermion line of $H_1$
in the Glauber region
\begin{eqnarray}
\frac{2p_1^+}{(p_1-k_1+l)^2-m_q^2+i\epsilon}\approx\frac{p_1^+}{p_1\cdot
l+i\epsilon}=\frac{1}{l^-+i\epsilon},\label{eis}
\end{eqnarray}
with the numerator $2p_1^+$ coming from Eq.~(\ref{trace}). The
denominator $[(k_1-l)^2-m_q^2](l^2-m_g^2)$ becomes $(|{\bf l}_T-{\bf
k}_{1T}|^2+m_q^2)(l_T^2+m_g^2)$ after the integrations over $l^+$
and then over $l^-$ in Eq.~(\ref{ssa1}). Due to the existence of
$l^\beta$ in Eq.~(\ref{trace}), only the region of $l_T\to k_{1T}$
generates a residual infrared divergence at NLO.

The rest of the procedure for factorizing the residual infrared
divergence is subtler than for the unpolarized hadron
hadroproduction. The LO polarized TMD parton density vanishes with
the fermion trace,
\begin{eqnarray}
\frac{1}{2}Tr[(\not p_1+m_H)\gamma_5\not s (\not p_1-\not k_1+m_q)] =0.
\label{trace0}
\end{eqnarray}
However, it is still legitimate to associate the factor
$l^\beta/(|{\bf l}_T-{\bf k}_{1T}|^2+m_q^2)$ with the LO polarized
TMD parton density, since Eq.~(\ref{trace0}) can be regarded as
resulting from the absence of $l^\beta$ in the case of no Glauber
gluon. The Fourier transformation of Eq.~(\ref{ssa1}) under the
eikonal approximation in Eq.~(\ref{eis}) leads to
\begin{eqnarray}
S_{L}^{(1)}&\approx&\int d^2b_l d^2b_rd^2b'_ld^2b'_r [-iS({\bf
b}_l)]d\hat\sigma^{(0)}_{i+j\to k+l}({\bf b}_l-{\bf b}'_l,{\bf
b}_r-{\bf b}'_r)e^{i{\bf k}_{1T}\cdot({\bf b}_l-{\bf b}_r)}e^{-i{\bf
k}_{3T}\cdot({\bf b}_l-{\bf b}_r-{\bf b}'_l+{\bf b}'_r)}
\nonumber\\
& &\times (m_H+m_q)\epsilon_{\alpha\beta}
s^\alpha\left(\frac{\partial}{\partial {\bf
b}'_{l\beta}}+\frac{\partial}{\partial {\bf b}'_{r\beta}}\right)\hat
f_{i/1}^{(0)}({\bf b}'_l,{\bf b}'_r)\cdots,\label{lef}
\end{eqnarray}
with the corresponding LO parton-level differential cross section
$d\sigma_{i+j\to k+l}^{(0)}$. We have written the LO polarized TMD
parton density in terms of the unpolarized one in Eq.~(\ref{LO}).
The derivatives with respect to ${\bf b}'_{l\beta}$ and ${\bf
b}'_{r\beta}$ correspond to $(l^\beta-k_1^\beta)$ and $k_1^\beta$ in
the transverse momentum space, respectively, whose sum gives
$l^\beta$ in Eq.~(\ref{trace}).

It is easy to verify that the LO cross section for the SSA
diminishes in our formalism. Without the Glauber gluon, we drop
$[-iS({\bf b}_l)]$ in Eq.~(\ref{lef}), obtaining
\begin{eqnarray}
S_{L}^{(0)}&\approx&\int d^2b_l d^2b_r d\hat\sigma^{(0)}_{i+j\to
k+l}({\bf b}_l,{\bf b}_r)e^{i({\bf k}_{1T}-{\bf k}_{3T})\cdot({\bf
b}_l-{\bf b}_r)}
\nonumber\\
& &\times (m_H+m_q)\epsilon_{\alpha\beta} s^\alpha\int
d^2b'_ld^2b'_re^{i{\bf k}_{1T}\cdot({\bf b}'_l-{\bf
b}'_r)}\left(\frac{\partial}{\partial {\bf
b}'_{l\beta}}+\frac{\partial}{\partial {\bf b}'_{r\beta}}\right)\hat
f_{i/1}^{(0)}({\bf b}'_l,{\bf b}'_r)\cdots,\label{lef1}
\end{eqnarray}
where the trivial variable changes ${\bf b}_l-{\bf b}'_l\to {\bf
b}_l$ and ${\bf b}_r-{\bf b}'_r\to {\bf b}_r$ have been made. A
trivial integration by parts then shows that the second line in the
above expression vanishes like ${\bf k}_{1T}^\beta-{\bf
k}_{1T}^\beta=0$.

At NNLO, the factorization for $S_{LR}^{(2)}$ with one Glauber gluon
on each side of the final-state cut follows the same steps as for
$S_{L}^{(1)}$. For the diagram with two Glauber gluons before the
final-state cut, we have the fermion trace
\begin{eqnarray}
& &\frac{1}{2}Tr[(\not p_1+m_H)\gamma_5\not s(\not p_1-\not k_1+\not
l_1+\not l_2+m_q)\gamma^+(\not p_1-\not k_1+\not
l_2+m_q)\gamma^+(\not p_1-\not k_1+m_q)]\nonumber\\
& & \approx i(m_H+m_q)\epsilon_{\alpha\beta}s^\alpha
(l_1^\beta+l_2^\beta) (2p_1^+)^2.\label{tra}
\end{eqnarray}
Equation~(\ref{tra}), exhibiting a pattern similar to
Eq.~(\ref{trace}), also hints that the factor
$(l_1^\beta+l_2^\beta)/(|{\bf l}_{1T}+{\bf l}_{2T}-{\bf
k}_{1T}|^2+m_q^2)$ is absorbed into the LO polarized TMD parton
density, and that $(2p_1^+)^2$ are employed in the eikonal
approximation of the two spectator propagators. The corresponding
factorization formula in the impact parameter space is then written
as
\begin{eqnarray}
S_{LL}^{(2)}&\approx&\int d^2b_ld^2b_rd^2b'_ld^2b'_r
\frac{1}{2}[-iS({\bf b}_l)]^2d\hat\sigma^{(0)}_{i+j\to k+l}({\bf
b}_l-{\bf b}'_l,{\bf b}_r-{\bf b}'_r)e^{i{\bf k}_{1T}\cdot({\bf
b}_l-{\bf b}_r)}e^{-i{\bf k}_{3T}\cdot({\bf b}_l-{\bf b}_r-{\bf
b}'_l+{\bf b}'_r)}\nonumber\\
& &\times (m_H+m_q)\epsilon_{\alpha\beta}
s^\alpha\left(\frac{\partial}{\partial {\bf
b}'_{l\beta}}+\frac{\partial}{\partial {\bf b}'_{r\beta}}\right)\hat
f_{i/1}^{(0)}({\bf b}'_l,{\bf b}'_r)\cdots.\label{2bs}
\end{eqnarray}
The above observation applies to higher loops trivially in the
simple toy model considered here.

We next include the collinear gluon exchanges of the type in
Fig.~\ref{fig1}(c). When a Glauber gluon carries the momentum $l_1$
and an ordinary collinear gluon carries $l_2$, the fermion trace in
Eq.~(\ref{tra}) still holds. The terms proportional to $l_1^\beta$
and $l_2^\beta$, after being integrated over $l_2$, lead to
$f_{i/1}^{(1)}$ multiplied by $l_1^\beta$, and the original
transverse-spin-dependent TMD parton density for the SSA with the
structure $(k_1-l_1)^\beta$, respectively. If the collinear gluon is
exchanged after the final-state cut, we obtain $f_{i/1}^{(1)}$
multiplied by $l_1^\beta$, and the transverse-spin-dependent TMD
parton density with the structure $k_1^\beta$, respectively. Since
diagrams for the TMD parton density are symmetric with respect to
the final-state cut, the sum of the NNLO diagrams with one collinear
gluon, and one Glauber gluon before the final-state cut gives
\begin{eqnarray}
l_1^\beta f_{i/1}^{(1)}({\bf k}_{1T}-{\bf l}_{1T},{\bf
k}_{1T})+\frac{1}{2}\left[(k_1-l_1)^\beta+ k_1^\beta\right]
f_{i/1}^{\perp(1)}({\bf k}_{1T}-{\bf l}_{1T},{\bf k}_{1T}),
\end{eqnarray}
in which $f_{i/1}^{\perp(1)}$ denotes the original
transverse-spin-dependent TMD parton density for the SSA.

Extending the above procedure to all orders, we derive the
factorization formula with Glauber gluons being emitted by the
spectator of $H_1$ and attaching to the active partons of $H_2$ and
$H_4$
\begin{eqnarray}
S&\approx&\int d^2b_ld^2b_rd^2b'_ld^2b'_r e^{-iS({\bf
b}_l)}d\hat\sigma_{i+j\to k+l}({\bf b}_l-{\bf b}'_l,{\bf b}_r-{\bf
b}'_r)e^{iS({\bf b}_r)}e^{i{\bf k}_{1T}\cdot({\bf b}_l-{\bf
b}_r)}e^{-i{\bf k}_{3T}\cdot({\bf b}_l-{\bf b}_r-{\bf b}'_l+{\bf
b}'_r)} \nonumber\\
& &\times (m_H+m_q)\epsilon_{\alpha\beta}
s^\alpha\left[\left(\frac{\partial}{\partial {\bf
b}'_{l\beta}}+\frac{\partial}{\partial {\bf b}'_{r\beta}}\right)\hat
f_{i/1}({\bf b}'_l,{\bf
b}'_r)-\frac{1}{2}\left(\frac{\partial}{\partial {\bf
b}'_{l\beta}}-\frac{\partial}{\partial {\bf b}'_{r\beta}}\right)\hat
f_{i/1}^\perp({\bf b}'_l,{\bf b}'_r)+\right]\cdots.\label{ssig1}
\end{eqnarray}
It is observed that the Glauber gluons do not break the universality
of the transverse-spin-dependent TMD parton density, and that the
Glauber factor extracted from the SSA is identical to the one from
the unpolarized hadron hadroproduction. If the Glauber factor is
absent, Eq.~(\ref{ssig1}) will reduce to the standard $k_T$
factorization formula
\begin{eqnarray}
& &\int d^2b_l d^2b_r d\hat\sigma_{i+j\to k+l}({\bf b}_l,{\bf
b}_r)e^{i({\bf k}_{1T}-{\bf k}_{3T})\cdot({\bf b}_l-{\bf b}_r)}
\nonumber\\
& &\times (m_H+m_q)\epsilon_{\alpha\beta} s^\alpha k_1^\beta\int
d^2b'_ld^2b'_re^{i{\bf k}_{1T}\cdot({\bf b}'_l-{\bf b}'_r)}\hat
f_{i/1}^\perp({\bf b}'_l,{\bf b}'_r)\cdots,\label{ssig2}
\end{eqnarray}
where the derivatives with respect to ${\bf b}'_{l\beta}$ and ${\bf
b}'_{r\beta}$ have applied to the Fourier factor to generate
$k_1^\beta$ through integration by parts. The factorization of the
corresponding transverse-spin-dependent TMD parton density in QCD
from polarized hadron hadroproduction, namely, the Sivers function,
will be studied in the same framework.

\section{CONCLUSION}

In this paper we have proposed the restoration of the $k_T$
factorization theorem for the hadron hadroproduction at low $p_T$ in
a simple toy model. The idea relies on the large separation of the
two relevant poles in different half planes of a loop momentum at
low $p_T$. The contour of a loop momentum is then deformed away from
the Glauber region, and the eikonalization holds for factorizing the
residual infrared divergence from the hadron-hadron collision.
Therefore, the universality of a TMD parton density is recovered at
the price that the $k_T$ factorization formula involves an
additional nonperturbative Glauber factor. It has been shown that
the Glauber factor is universal: for example, the same Glauber
factor has been extracted from the unpolarized and polarized hadron
hadroproduction. Hence, its behavior can be constrained
experimentally from some processes, and then employed to make
predictions for others. Our observation also applies to the $W$
boson plus jet production and to the direct photon production, for
which the momentum $k_2$ ($k_3$) is carried by a gluonic parton
(gauge bosons). In these processes the $\delta$-function leading to
the Glauber divergence appears after summing over the attachments to
the gluonic parton, the quark carrying the momentum $k_4$, and the
virtual quark. The same divergence has been identified in the
color-suppressed tree amplitudes \cite{LM09} of two-body nonleptonic
$B$ meson decays in the perturbative QCD (PQCD) approach, which is
based on the $k_T$ factorization theorem \cite{LY1,CL,YL,KLS,LUY}. A
Glauber factor has been introduced into the PQCD formulas, which
enhances the color-suppressed tree amplitudes significantly, such
that the known $\pi\pi$ and $\pi K$ puzzles were resolved
\cite{LM09}.

In a forthcoming paper we shall discuss the $k_T$ factorization of
the Glauber divergence from low-$p_T$ hadron hadroproduction in real
QCD, where the spectator lines in Fig.~\ref{fig1} are replaced by
infinitely many rung gluons, forming the so-called ladder diagrams.
Moreover, gluonic partons, instead of quark partons, play a major
role in the small-$x$ region. It has been known that the region with
strong rapidity ordering for the ladder diagrams gives a dominant
contribution, which has been summed into the
Balitsky-Fadin-Kuraev-Lipatov evolution equation \cite{BFKL}. The
strong rapidity ordering corresponds to $k_1^+\ll p_1^+$ in the
present toy model, under which the two poles in Eq.~(\ref{pole}) are
far apart from each other, so the eikonal approximation should also
hold for real QCD processes. A difference arises from additional
color degrees of freedom of quarks and gluons. Equation~(\ref{eik1})
is then a consequence of the summation over the attachments of the
radiative gluon to the active parton lines in $H_2$ and $H_4$ and to
the hard gluon line. To employ Eq.~(\ref{eiko}) in QCD, the diagrams
with triple gluon vertices should be included too. In this case the
eikonalization and the Ward identity are needed in order to work out
the summation over the attachments of the Glauber gluons to all rung
gluons.

We noticed that our results for the unpolarized hadron
hadroproduction in this work (with the preprint number
arXiv:0904.4150) have been confirmed by a later publication
\cite{XY10} with the following one-to-one correspondence: the
exponent $G(R_\perp)$ in Eq.~(7) of \cite{XY10} corresponds to
$S({\bf b}_l)$ in our Eq.~(\ref{left}). The explicit expression
$G(R_\perp)=K_0(\lambda R_\perp)/(2\pi)$ in \cite{XY10} is identical
to our Eq.~(\ref{sf}), where the infrared regulator was chosen as
$m_g$. The difference is that we did not distinguish the coupling
constants $g_1$ for the gluon attachments to the lower parton line,
from $g_2$ for the attachments to the upper parton line. The
concluding equation~(8) in \cite{XY10} is consistent with our
Eq.~(\ref{sig1}) with the correspondence between
$\exp\{-igg_2[G(R_\perp)-G(R'_\perp)]\}$ and $\exp\{-i[S({\bf
b}_l)-S({\bf b}_r)]\}$ for the Glauber factor. Though both groups
have factorized the Glauber divergences, the interpretations of the
final result are opposite. We conclude that the universality of a
TMD parton density has been restored, but they do not, because they
have regarded the Glauber factor as part of a TMD parton density.
Actually, the Glauber factor, having been factorized, should be
treated as an independent input in the $k_T$ factorization theorem,
and can be constrained experimentally or derived by nonperturbative
methods.

\vskip 1.0cm We thank J.W. Qiu, W. Vogelsang, C.P. Yuan, and F. Yuan
for useful discussions. This work was supported by the National
Science Council of R.O.C. under the Grant No.
NSC-98-2112-M-001-015-MY3, and by the National Center for
Theoretical Sciences of R.O.C..

\end{document}